\documentclass[12pt,epsfig]{article}
\newcommand{\ba}{\begin{array}{c}}
\newcommand{\baz}{\begin{array}{cc}}
\newcommand{\bad}{\begin{array}{ccc}}
\newcommand{\bav}{\begin{array}{cccc}}
\newcommand{\ea}{\end{array}}

\usepackage{amssymb} 
\usepackage{epsfig}
\usepackage{float}
\newcommand{\be}{\begin{equation}}
\newcommand{\ee}{\end{equation}}
\newcommand{\bea}{\begin{eqnarray}}
\newcommand{\eea}{\end{eqnarray}}
\begin{document}

\begin{center}
\bf {A comment on the lepton mixing matrix}
\end{center}

\begin{center}
C. Jarlskog 
\end{center}

\begin{center}
{\em Division of Mathematical Physics\\
LTH, Lund University\\
Box 118, S-22100 Lund, Sweden}
\end{center}

\begin{abstract}

A recent parameterisation scheme developed for the quark mixing matrix
is shown to be easily applicable to the lepton mixing matrix as well.

\end{abstract}

\section{The parameterisation}

Recently, guided by experiments, an explicit form was suggested
\cite{cejacab} for
the quark mixing matrix to emphasise its Cabibbo substructure.
In this parameterisation, the smallest angle, denoted by $\theta_3$ 
is of order $\lambda^2$, 
where $\lambda$ is the parameter introduced by Wolfenstein in
his empirical scheme \cite{lincoln}. An essential feature is that
one may easily choose any arbitrary
angle of the experimentally studied unitarity triangle to be
one of the parameters of the matrix. Evidently, one could
choose any angle of any of the other triangles as well. 

The purpose of this note is to show how, within this framework, the currently
known lepton mixing matrix is reproduced.  

The idea in Ref.\cite{cejacab} was to write the mixing matrix in 
the exact form
\begin{equation} V = V_0 + s_3 V_1 + (1-c_3) V_2
\label{defv} 
\end{equation}
where $s_3 = sin\theta_3$, $c_3 = cos \theta_3$ and the matrices 
$V_j$, $j=0-2$, are given by
\begin{eqnarray} 
V_0 &=&
\left( \begin{array}{cc} U & 
\begin{array}{c} 0 \\ 0 \end{array} \\ \begin{array}{cc}
0~& 0 \end{array} & 1 
\end{array}
\right)  \\
 V_1 &=&
\left( \begin{array}{ccc}
0& 0 & a_1 \\
0 & 0 & a_2\\
b_1^\star & b_2^\star & 0
\end{array}
\right) \equiv \left( \begin{array}{cc} 0 & \vert A> \\ <B \vert & 0 
\end{array} \right)\\
 V_2 &=&
\left( \begin{array}{cc} \vert A > < B \vert & 
\begin{array}{c} 0 \\ 0 \end{array} \\ \begin{array}{cc}
0~& 0 \end{array} & -1 
\end{array}
\right) 
\end{eqnarray}
Here $U$ is a two-by-two unitary matrix and
\begin{equation}
\vert A >= \left( \begin{array}{c} a_1 \\ a_2 \end{array} \right), ~~~~
\vert B >= \left( \begin{array}{c} b_1 \\ b_2 \end{array} \right)
\end{equation}
and $(\vert A > < B \vert)_{ij} \equiv a_i b_j^\star$. Furthermore,
vector $A$ is normalised to one and the two vectors $A$ and $B$ are
related to one another,  
$\vert B> \equiv -U^\dagger \vert A>$. Thus we have
\begin{equation}
<A \vert A> =<B \vert B> =1, ~~\vert A> = -U \vert B>,
 ~~~\vert B> = -U^\dagger \vert A>
 \label{relab}
\end{equation}
It is a trivial task to check that $V$ is unitary, as was
described in Ref.\cite{cejacab}.
For the case of quarks, the obvious choice
for $U$ is to take it to be
a rotation matrix, which was denoted by $R_2(\Phi)$ in Ref.\cite{cejacab}.
Thereby the Cabibbo substructure is manifestly exhibited.
Finally, the CP violation parameter $J$ defined by
\begin{equation}
Im (V_{\alpha j}V_{\beta k}V^\star_{\alpha k}V^\star_{\beta j})= J~\sum_{\gamma , l}^{}
\epsilon_{\alpha\beta\gamma} \epsilon_{jkl}
\label{defj}
\end{equation}
is in this parameterisation given by
\begin{equation}
J = s^2_3 c_3 sin\Phi cos\Phi Im(a_1^\star a_2) = 
s^2_3 c_3 sin\Phi cos\Phi Im(b_1^\star b_2)
\label{defja}
\end{equation}
We now turn to the case of leptons.

\section{Lepton mixing matrix}

The lepton mixing matrix is a product of two matrices. One of these
contains the so called Majorana phases and the other has the same
form as the quark mixing matrix. For simplicity, in the following 
we shall refer to the latter matrix as the lepton mixing matrix. 
The values of the matrix elements of this matrix 
may be found in the article by Kayser in the Review of Particle Physics
\cite{pdg}.
Kayser quotes
\begin{equation}
V \simeq \left[
\begin{array}{ccc}
c & s & s_{13}e^{-i\delta} \\
-s/ \sqrt{2} & c/ \sqrt{2} & 1 /\sqrt{2} \\
s / \sqrt{2} & - c / \sqrt{2} &1 / \sqrt{2}
\end{array}
\right]
\label{kayserv}
\end{equation}
To obtain this form in the above parameterisation we first note
that Eq.(\ref{defv}) yields
\begin{equation}
V_{33}= c_3.~~
V_{13} = a_1 s_3, ~~V_{23} = a_2 s_3, ~~
V_{31} = b^\star_1 s_3, ~~
V_{32} = b^\star_2 s_3
\end{equation}
Therefore we put $c_3 = {1 \over \sqrt{2}}$ and choose
\begin{equation}
U = R_2(\theta)= \left( \begin{array}{cc}
c& s \\ -s& c \end{array} \right)
\end{equation}
where $c$ and $s$ are the parameters appearing in Eq.(\ref{kayserv}). Next we put
\begin{equation}
\vert A > = \left( \begin{array}{c} 
\hat{s} e^{-i \delta} \\ \hat{c}
\end{array} \right) \simeq \left( \begin{array}{c}
\hat{\theta}e^{-i \delta} \\ 1
\end{array} \right)~+ O(\hat{\theta}^2)
\label{alep}
\end{equation}
From Eq.(\ref{relab}) follows that
\begin{equation}
\vert B > \simeq \left( \begin{array}{c}
s\\ -c \end{array} \right) - \hat{\theta}e^{-i \delta}
\left( \begin{array}{c} c  \\
s  \end{array} \right)~ + O(\hat{\theta}^2) 
\end{equation}
The two-by-two matrix in the upper left corner of
Eq.(\ref{defv}) is in the present case given by
\begin{equation}
\left( \begin{array}{cc} V_{11}& V_{12}\\
V_{21} & V_{22} \end{array} \right) = 
R_2(\theta)+ (1-c_3) \vert A> < B \vert
\end{equation}
A simple computation yields
\begin{equation}
\left( \begin{array}{cc} V_{11}& V_{12}\\
V_{21} & V_{22} \end{array} \right) = 
\left( \begin{array}{cc}
c& s \\ -sc_3& c c_3 \end{array} \right)+ (1-c_3)\hat{\theta}
\left( \begin{array}{cc}
 se^{-i \delta}& -ce^{-i \delta} \\ 
-c e^{i \delta}& -se^{i \delta} \end{array} \right)
~ + O(\hat{\theta}^2)
\end{equation}
Inserting these results into the matrix $V$ we find
\begin{equation}
V \simeq \left[
\begin{array}{ccc}
c & s & \hat{\theta} e^{-i\delta}/\sqrt{2} \\
-s/ \sqrt{2} & c/ \sqrt{2} & 1 /\sqrt{2} \\
s / \sqrt{2} & - c / \sqrt{2} &1 / \sqrt{2}
\end{array}
\right]
\label{cejav}
\end{equation}
where we have substituted $c_3 = {1 \over \sqrt{2}}$, and following Ref.\cite{pdg} 
we have systematically neglected terms of order 
$\hat{\theta}$ as compared to the leading terms. Identifying 
${\hat{\theta} \over \sqrt{2}}$ with $s_{13}$ we see that the
two matrices in Eqs.(\ref{kayserv}) and [\ref{cejav}] have exactly the same enteries.

Finally, remembering that $\Phi$ has been replaced by $\theta$ and $c_3 = {1 \over \sqrt{2}}$,
from Eqs.(\ref{defja}) and (\ref{alep}) we obtain
\begin{equation}
J \simeq {sc \over 2 \sqrt{2}}\hat{\theta} sin\delta 
= {sc \over 2 }s_{13}sin\delta 
\end{equation}
This quantity is not necessarily small and leaves us with the hope
that CP violation in neutrino oscillations may be observable.


\begin{thebibliography}{99}
\baselineskip=15pt

\bibitem{cejacab}
C. Jarlskog, hep-ph/0503199

\bibitem{lincoln} L. Wolfenstein, Phys. Rev. Lett. \underline{51} (1983) 1945 

\bibitem{pdg} B. Kayser {\it in} Review of Particle Physics, S. Eidelman et al.,
Phys. Lett \underline {B592} (2004) 1


\end{thebibliography}
\end{document}